\documentstyle[12pt,amsmath,latexsym]{article}
\def\Box{\hbox{$\sqcup$\kern-0.66em\lower0.03ex\hbox{$\sqcap$}}}
\begin{document}
\begin{titlepage}
\begin{flushright}
IFUP--TH 39/99 \\
LPTENS--99/24
\end{flushright}
\vskip 1truecm
\begin{center}
\Large\bf
ADM approach to 2+1 dimensional gravity\\ coupled to particles
\footnote{This work is  supported in part
  by M.U.R.S.T.}.
\end{center}
\vskip 1truecm
\begin{center}
{Pietro Menotti} \\ 
{\small\it Dipartimento di Fisica dell'Universit{\`a}, Pisa 56100, 
Italy and}\\
{\small\it INFN, Sezione di Pisa}\\
\end{center}
\vskip .8truecm
\begin{center}
{Domenico Seminara\footnote{CEE Post-doctoral Fellow under contract
FMRX CT96-0045 }} \\  
{\small\it Laboratoire de Physique Th{\'e}orique, 
{\'E}cole Normale Sup{\'e}rieure\footnote{Unit{\'e} Mixte associee au
Centre de la Recherche Scientifique et  \'a l' \'Ecole Normale
Sup\'erieure.}.\\
F-75231, Paris CEDEX 05, France}\\
\end{center}
\begin{center}
July 1999
\end{center}
\end{titlepage} 
\begin{abstract}
We develop the canonical ADM approach to 2+1 dimensional gravity in presence 
of point particles. The instantaneous York gauge can
be applied for open  
universes or universes with the topology of the sphere. The sequence 
of canonical ADM equations is solved in terms of the conformal
factor. A simple derivation is given for the solution of the two body
problem. 
A geometrical characterization is given for the apparent singularities
occurring in the N-body problem and it is shown how the Garnier
hamiltonian system arises in the ADM treatment by considering the time
development of the conformal factor at the locations where the
extrinsic curvature tensor vanishes.
\end{abstract}
\section{Introduction}
\label{introd}
After the seminal paper by Deser Jackiw and 't Hooft \cite{DJH} the
problem of supplying a solution to $2+1$ dimensional gravity has been
approached from different viewpoints and with different
techniques. Moncrief \cite{moncrief} and Hosoya and Nakao
\cite{hosoya} gave the hamiltonian treatment in
absence of particles. In \cite{moncrief,hosoya} the complete reduction of the
hamiltonian to the physical parameters i.e. the moduli of the time
slices was obtained in the York (minimal surface) gauge; the problem
can be dealt 
with explicitly for the torus while for higher genus, even if well
defined, is far from trivial. The solution of \cite{moncrief} and
\cite{hosoya} were exploited in \cite{carlip} and  \cite{hosoya2} for the
quantization of the theory.

In presence of particles progress was made in the papers by
Bellini, Ciafaloni and Valtancoli \cite{BCV} and by Welling
\cite{welling} in the first order 
formalism by going over to the instantaneous gauge; in these papers it
was shown that the problem is equivalent to the solution of a 
Riemann- Hilbert problem. It can be solved explicitly for the case of two
particles, in terms of hypergeometric functions. For three or more
particles one encounters a feature known in the mathematical literature
as apparent singularities. Such apparent singularities evolve
according to a hamiltonian system of equations known as the Garnier
system \cite{yoshida} which is derived in the mathematical literature
from the isomonodromicity condition.

A different approach was put forward by 't Hooft \cite{hooft} by
describing the evolving Cauchy surfaces in terms of polygonal tiles
which join along segments where the extrinsic curvature is
singular. The dynamics of the system is codified in transition rules
which intervene when e.g. the length of a side of a polygon goes to
zero or when a particle collides with the side of a polygon. A similar
but different approach was given by  Waelbroek
\cite{wael}. Quantization schemes were given in
\cite{hooft2,waelbroek2,matschull}.

In this paper we shall consider $2+1$ dimensional gravity in presence
of massive particles by exploiting the hamiltonian formulation. Most
techniques apply to massless particles as well, but here we shall
not be concerned with this case.
Thus our setting
is a second order formalism and the basic equations are those derived
systematically from the variation of the ADM action. The reason
for such a development is to give a treatment which resides completely
within the canonical framework. We shall see that all results obtained
in \cite{BCV} \cite{welling} come out in simple fashion from such an
approach; moreover we shall prove that the Garnier hamiltonian system
is a direct outcome of the canonical ADM formalism. Again the
exploitation of the instantaneous York gauge plays a major role; the
technical advantage of such a gauge is to reduce an equation of
sinh-Gordon to one of Liouville type to which
powerful methods of complex analysis apply.  Unfortunately the
applicability of such a gauge is restricted to universes of spherical
topology or to open ones.

We shall see that such an approach provides an elementary way to solve
the two body problem: the exact solution of the motion can be derived even
before the explicit computation of the metric, which combined with the
qualitative knowledge to the asymptotic metric gives a complete
description of the scattering. The exact metric is obtained by solving
a Liouville equation, leading of course to the same results obtained
in \cite{BCV} and \cite{welling}. 

One of the aims of presenting a canonical solution of the problem is to
provide a framework for the quantization of the theory but here we
shall be concerned solely with the classical theory.

The plan of the paper is the following: After recalling in sect.2 
the ADM canonical formulation of gravity coupled
with point particles we proceed in sect.3 to the solution of the
canonical equations. These will be solved in the following order:
First one computes the momenta canonically conjugate to the space
metric; they turn out to be rational analytic or antianalytic
functions of the complex coordinates of the plane. The knowledge of
such canonical momenta allows one to write down the partial
differential equation for the conformal factor or better for a more
fundamental quantity which we shall call the reduced conformal
factor. This is a Liouville equation whose sources are located at the
particle position; in addition when the number of particles ${\cal N}$
is greater than $2$ new sources appear. These are located at the
points where the extrinsic curvature tensor vanishes and are identified
as the position of the apparent singularities of the fuchsian
differential equation which underlies the expression of the conformal
factor. In the ADM equation for the conformal factor $\sigma$ the
sources appear as function of the conformal factor itself. On the other
hand it is shown in the text that due to the vanishing of the inverse $g^{ij}$
of the spacial metric on the particles,
the sources simplify 
and are just given by 
the particle masses. The Liouville equation admits a
whole family of solutions parametrized by a number $M$ whose values
runs from zero to $4\pi$ and it is related to
the asymptotic behavior of the conformal factor; it is nothing else
than the total energy of the system in adimensional units. 
A byproduct of the ADM equation for
the time derivative of $\sigma$ is that such parameter $M$ is constant
i.e. the energy is conserved.

The Lagrange multiplier $N$ which appears in the ADM
metric in the combination $-N^2 dt^2$ satisfies a differential
equation, which again due to the vanishing of $g^{ij}$ on the
particles reduces to a homogeneous linear differential equation. This
is shown to be simply solved by the derivative of the conformal factor
with respect to the parameter $M$. An elementary theorem proves that
such a solution is unique. The final ingredient of the ADM metric
i.e. the shift functions $N^i$ are expressed in terms of the spacial
derivative of $N$.

In sect.4 we write the Hamilton equation for the time variation of the
particle positions and momenta. These are 
explicited for the general ${\cal N}$-body problem in terms of the
coefficient of the fuchsian equation which underly the conformal
factor. In the two body case, which is dealt with in sect.5, they give
rise to an elementary system of two 
differential equations for the particle positions and momenta, which
solved provides the particle trajectories. The exact metric can be
given in term of the classical solution of the Liouville equation. In
sect.6 we give the general form of the asymptotic metric in the York
instantaneous gauge. 

In sect.7  we give the treatment of the problem for ${\cal N}>2$. It
is shown how the fuchsian nature of the differential equation is related
to the conservation of momenta and to  a generalized conservation law
which appears related to the transformation of the action under space
dilatations. 

As mentioned above, with more than two particles there are 
apparent singularities and one has to provide evolution equations for
their positions and residues. We show that all this information in
contained in the ADM equation for the the time derivative of the
conformal factor. In fact by exploiting Schwarz's relation between the
coefficient of the fuchsian differential equation and the reduced
conformal factor one obtains the Garnier equation by equating the
residues at the polar singularities at the position of the apparent
singularities.

\section{The action}

The action for the gravitational field in any dimension, including the
boundary terms is given by \cite{hawkinghunter}
$$
S_{Grav}=\frac{c^3}{16\pi
G_N}\int_{{\cal M}}d^{(D+1)}x \sqrt{\rm g}~ {\cal R} + \frac{c^3}{8\pi
G_N}\int_{\Sigma_0}^{\Sigma_1}d^Dx \sqrt{g}\,K + 
$$
\begin{equation} 
+\frac{c^3}{8\pi
G_N}\int_B d^Dx \sqrt{-\gamma}\,\Theta + \frac{c^3}{8\pi
G_N}\int_{B_0}^{B_1}d^{D-1}x \sqrt{\sigma}~ \eta 
\end{equation}
where $\rm g_{\mu \nu}$, ${\cal R}$
are the $D+1$ dimensional metric and curvature, $g_{ij}$ the
$D$-dimensional metric of the constant time slices $\Sigma_t$;
$\Sigma_0$ and $\Sigma_1$ the initial and final time slices; $K_{ij}$
the second 
fundamental form of the time slices $\Sigma_t$ and $\Theta_{ij}$ the second
fundamental form of the lateral boundary $B$ whose volume form is
$\sqrt{-\gamma}$; $n^\mu $ is the future
pointing unit  
normal to the time slices and $u^\mu $ the outward pointing unit
normal to $B$; $B_t= \Sigma_t\cap B$, $\sqrt{\sigma}$ is the volume
form induced on $B_t$ and $\sinh\eta=n_\mu u^\mu $. 
In presence of particles we must add to the
gravitational action the matter term 
\begin{equation} S_m =-\int
dt \sum_n {\rm m}_n c\sqrt{- {\rm g}_{\mu \nu}(q_n)\dot q_n^\mu \dot
q_n^\nu} 
\end{equation} 
where we have chosen the gauge $q_n^0=t$.  At the classical level it
is convenient to multiply the action by $\frac{16 \pi G_N}{c^3}$; the
matter action e.g. becomes 
\begin{equation} S_m =-\int dt
\sum_n m_n\sqrt{-{\rm g}_{\mu \nu}(q_n)\dot q_n^\mu \dot q_n^\nu}
\end{equation} 
with $m_n = \frac{16 \pi G_N {\rm m}_n}{c^2}$; it is
a real number $0<m_n<4\pi$.  In the ADM notation where the metric is
written as \cite{adm} 
\begin{equation} ds^2 = - N^2 dt^2
+g_{ij}(dx^i+N^idt)(dx^j+N^jdt) 
\end{equation} 
the volume term can be
rewritten as \cite{adm}
\begin{equation} \label{eq:1ADMSEH} 
S_{H}= \int_{{\cal
M}}d^Dxdt\ N\sqrt{g}\big[ R+K_{ij}K^{ij}
-K^2+2(n^\mu K-a^\mu )_{;\,\mu }\big], 
\end{equation} 
where $R$ is
the $D$ dimensional curvature related to $g_{ij}$ and $a^\mu $ is the
acceleration vector 
$a^\mu = n^\nu n^\mu _{~;\nu}$.  
Going over to the canonical variables 
\begin{equation} P_{ni} =
\frac{\partial{\cal L}}{\partial \dot q_n^i}= m_n \frac{{\rm
g}_{i\nu}(q_n) \dot q_n^\nu}{\sqrt{-{\rm g}_{\rho\sigma}(q_n)\dot
q_n^\rho \dot q_n^\sigma}}= m_n \frac{g_{ij}(q_n) (\dot q_n^j
+N^j)}{\sqrt{N^2 - g_{ij}(\dot q_n^i + N^i )(\dot q_n^j +N^j)}}
\end{equation} 
the matter action can be rewritten as 
\begin{equation} 
S_m=\int\!d t \sum_n\Big(P_{ni}\, \dot
q_n^i+N^i(q_n) P_{ni} - N(q_n) \sqrt{P_{ni} P_{nj} g^{ij}(q_n)+
m_n^2}\Big). 
\end{equation}
On the other hand the gravitational action expressed in terms of the
canonical variables becomes \cite{hawkinghunter,wald,MTW}
\begin{multline} 
S_{Grav} = \int_{{\cal M}} dt d^Dx
\left[ \pi^{ij}\dot g_{ij} - N^i H_i - NH\right]+ \\
+2\int dt \int_{B_t} d^{(D-1)}x \,\sqrt{\sigma} N 
\left( K_{B_t}+\frac{\eta}{\cosh\eta}\nabla_\alpha v^\alpha\right)
-2\int dt \int_{B_t} d^{(D-1)}x\, r_\alpha
\pi^{\alpha\beta}_{(\sigma)} N_\beta. 
\end{multline} 
The symbol
$K_{B_t}$ stands for the extrinsic curvature of $B_t$ as a
surface embedded in 
$\Sigma_t$, $v^\alpha\equiv \displaystyle{\frac{1}{\cosh\eta}\left 
(n^\alpha-\sinh\eta~ u^\alpha\right)}$ and $r_\alpha$ is the versor normal to
$B_t$ in $\Sigma_t$.    
The subscript $\sigma$ in $\pi^{\alpha\beta}_{(\sigma)}$ is a reminder 
that it has to be considered a tensor density with respect to
the measure $\sqrt{\sigma}$. 

\section{The field equations} 

\subsection{The conjugate momenta} 

Variation of the action with respect to $N^i$ gives
\begin{equation}\label{piz}
H_i=-2\sqrt{g}\nabla_{j}\frac{\pi^j_{~i}}{\sqrt{g}}-\sum_n\delta^2 (
x-q_n)P_{ni} =0 
\end{equation}
where $\nabla$ is the covariant derivative with respect to the induced
metric $g_{ij}$.
In a York $K={\rm const.}$ gauge we have 
\begin{equation}\label{pizb}
2\sqrt{g}\nabla_{j}\frac{\pi^{Tj}_{~i}}{\sqrt{g}}=-\sum_n\delta^2
(x- q_n)P_{ni} 
\end{equation} 
where $\pi^T$ is the
traceless part of $\pi^{ij}$. When $K$ vanishes, $\pi^{ij}$ becomes
traceless since in $2+1$ dimensions
\begin{equation}\label{conjmom} 
\pi^{ij}\equiv \sqrt{g}(K^{ij} - g^{ij} K) 
\end{equation} 
and thus $\pi^i_{~j}$ has
only two independent components. Using the complex coordinates $z= x
+i y$ and $\bar z =x-iy$, eq.(\ref{pizb}) reduces to 
\begin{equation} 
\partial_{\bar z}\pi^{\bar z}_{~z} =
-\frac{1}{2}\sum_n P_{n z} \delta^2(z- z_n) 
\end{equation}
\begin{equation} 
\partial_{z}\pi^z_{~\bar z} = -\frac{1}{2}\sum_n P_{n
\bar z} \delta^2(z - z_n) 
\end{equation} 
with $\pi^{\bar
z}_{~z} = \pi^x_{~x} - i~ \pi^x_{~y}$ and $\int \delta^2(z) dx dy =1$.
They are solved by 
\begin{equation}\label{solpiz}
\pi^{\bar
z}_{~z}=-\frac{1}{2\pi}\sum_n \frac{P_{nz}}{z - z_n} 
\end{equation}
\begin{equation}\label{solpizb}
\pi^{z}_{~\bar z} = -\frac{1}{2\pi}\sum_n \frac{P_{n
\bar z}} {\bar z - \bar z_n}.  
\end{equation}
We shall work in the c.m. frame i.e. with $\sum_n P_n=0$. With such a
restriction $\pi^{\bar z}_{~z}$ and $\pi^{z}_{~\bar z}$ decrease at
infinity at least like $1/|z|^2$. In principle one could add to the
solution (\ref{solpiz}) (and (\ref{solpizb})) and arbitrary analytic
(antianalytic) function. Our choice to maintain the behavior
$\pi^a_{~b} = O(1/|z|^2)$ will give rise, as discussed in sect.3.3 to
a well defined asymptotic behavior of the conformal factor
$e^{2\sigma} \sim (z\bar z)^{-\mu }$ and $N\sim \ln(z\bar z)$. Thus the
choice of solutions (\ref{solpiz},\ref{solpizb}) is a way to impose
the form of the asymptotic metric.

As 
we already mentioned in the introduction and will be shown below,
one can apply the instantaneous York gauge $K=0$ only to the topology
of the plane (open universes) or the topology of the sphere.   
The sphere can be described by the metric
\begin{equation} 
g_{ij} = \frac{8 e^{2\sigma}}{(1+z\bar z)^2}
\end{equation} 
where $\sigma$ is a conformal factor regular at infinity. The
equations for the $\pi^a_{~b}$ are still provided by
eq.(\ref{piz},\ref{pizb}) and their solution by
eq.(\ref{solpiz},\ref{solpizb}). On the other hand for the sphere
the point $z=\infty$ is a regular point. Imposition of the regularity of the 
tensor $g^{-1/2} \pi^a_{~b}$ at infinity, obtained through the
transformation $z' = 1/z$ imposes for $\pi^{\bar z}_{~z}$ the behavior
$\pi^{\bar z}_{~z} \approx z^{-n}$ with $n\geq 4$. This is reflected on
the following sum rules on the momenta $\sum_n P_{nz} =0$, $\sum_n
P_{nz}z_n =0$, $\sum_n P_{nz}z^2_n =0$. Thus except for the static case
in which all momenta vanish we need at least four particles on the
sphere. In the following we shall be mainly interested in the plane
topology.

The tracelessness of the momenta $\pi^{ij}$ in the $K=0$ gauge entails
some conservation laws which relate combinations of particle positions
and momenta with no reference to the field variables. We shall  see in
sect.7 that such relations can be derived directly form the equations
of motions. 

We want to examine here the expression of the conservation of angular
momentum in the $K=0$ gauge in the context of the hamiltonian formalism.
Under a rotation 
$$
q_n^i \rightarrow q_n^i + \varepsilon q_n^j\epsilon_j^{~i};~~~~
P_{ni} \rightarrow P_{ni} - \varepsilon \epsilon_i^{~j} P_{nj}; 
$$
\begin{equation}
g_{ij}(x^k) \rightarrow g_{ij}(x^k -\varepsilon x^l\epsilon_l^{~k})
-\varepsilon \epsilon_i^{~i'}g_{i'j}(x) -\varepsilon
\epsilon_j^{~j'}g_{ij'}(x) 
\end{equation}
and similar transformations on the remaining field variables, the
volume hamiltonian extended to a disk of radius $r$ and the boundary
hamiltonian computed on a circle of radius $r$ are left invariant. The
term $\sum_n P_{ni} \dot q_n^i$ is also left invariant while the term $\pi^{ij}
\dot g_{ij}$ is left identically zero. As a consequence the quantity
\begin{equation}
\sum_n x_n^i \epsilon_i^{~j} P_{nj} 
\end{equation}
is a constant of motion. This is the expression of the conservation of
angular momentum in the $K=0$ gauge. 
 
\subsection{The conformal factor}

The variation of the action under $N$ gives the hamiltonian constraint
\cite{wald,MTW} 
\begin{equation}\label{Nvariation}
H =\frac{1}{\sqrt g}\Big[\pi^a_{\ b}\pi^b_{\ a}-(\pi^c_{\
c})^2\Big]-\sqrt g R
+\sum_n\delta^2(z- z_n)\sqrt{m_n^2 +P_{ni}~g^{ij}P_{nj}}=0.  
\end{equation}
which in the instantaneous York gauge, i.e. $K=0$ becomes the non
linear equation 
\begin{equation}
\label{sigmaeq}
2\Delta\sigma=-\pi^a_{\ b}\pi^b_{\ a}e^{-2\sigma}-\sum_n
\delta^2(z- z_n)\sqrt{m_n^2 + 4 P_{nz} P_{n\bar z}~e^{-2\sigma}}. 
\end{equation}
Integrating eq.(\ref{Nvariation}) over all space and using the
Gauss-Bonnet theorem we see that, due to the positivity of
$\pi^a_{~b}\pi^b_{~a}$ the $K=0$ gauge is applicable only to closed  surfaces
of genus $0$ or to open universes \cite{welling}. We shall be concerned in the
following with open universes. 

The term $\pi^a_{~b}\pi^b_{~a} = 2 \pi^{\bar z}_{~z}\pi^z_{~\bar z}$ is the
product of an analytic $\pi^{\bar z}_{~z}(z)$, and an antianalytic
function and as such the laplacian of its logarithm is zero except on
the singularities.  
The structure of $\pi^{\bar z}_{~z}$, in the c.m. frame where $\sum_nP_n=0$, is
\begin{equation}
\pi^{\bar z}_{~z} = \frac{p_{{\cal N}-2}(z)}{\prod_n (z-z_n)}
\end{equation}
with $p_{{\cal N}-2}(z)$ a polynomial of degree ${\cal N}-2$, being
${\cal N}$ the
number of particles. It will possess ${\cal N}-2$ zeros in the complex plane
and this is the origin in the present approach of the so called
apparent singularities. Thus in the ADM approach  the apparent
singularities are given by the zeros of the extrinsic curvature tensor
$K_{ij}$ because we have the general relation eq.(\ref{conjmom})
and in our gauge $K=0$. Due to the Gauss-- Codazzi relations they are
also the points where the intrinsic curvature of the two dimensional
time slice vanish. We shall denote by $z_A$ the location of the
apparent singularities.
We shall now go over to the function 
\begin{equation}\label{sigmatilde}
2\tilde \sigma = 2 \sigma - \ln(2 \pi^{\bar z}_{~z}\pi^z_{~\bar z})
\end{equation}
and thus we have
\begin{equation}
\label{eqsigmatilde}
2\Delta\tilde\sigma=-e^{-2\tilde\sigma}-\sum_n
\delta^2(z- z_n)(\sqrt{m_n^2 + 4 P_{nz}P_{n\bar z} e^{-2\sigma}}-4\pi)-\sum_A
4\pi\delta^2(z- z_A)   
\end{equation}
which we want now to discuss near the particle singularities.

Near the $n$ particle, which to simplify the notation we suppose
to be placed at $z=0$, eq.(\ref{eqsigmatilde}) reduces to 
\begin{equation} 
\label{oneparticle} 2 \Delta \tilde \sigma = -
e^{-2\tilde\sigma}- 4\pi (a_n -1) \delta^2(z) 
\end{equation} 
with
\begin
{equation}
4 \pi a_n = \sqrt{m_n^2 + 4 ~C_n P_{nz} P_{n\bar z}} 
\end{equation}
and $C_n = e^{-2\sigma(0)}$. $C_n$
cannot be infinity otherwise $\tilde \sigma$ is not a solution of
eq.(\ref{oneparticle}). Eq.(\ref{oneparticle}) can be solved exactly
\cite{kra} by performing the transformation $w =
(z/\Lambda)^{a_n}$ on the solution of the Liouville equation $2\Delta
\tilde\sigma = - e^{-2\tilde\sigma}$ representing the Poincar\'e
pseudosphere, to obtain 
\begin{equation} e^{-2\tilde\sigma}=\frac{8 a_n^2}{\Lambda^2}
\frac{(z\bar z/\Lambda^2)^{a_n-1}}{\left(1-(z \bar
z/\Lambda^2)^{a_n}\right)^2} 
\end{equation} 
with arbitrary $\Lambda$.  In particular, for $|z|\rightarrow 0$ we have 
\begin{equation} e^{-2\tilde\sigma}\sim \frac{8a^2_n}{\Lambda^2}
\left(\frac{z\bar z}{\Lambda^2}\right)^{a_n-1}. 
\end{equation} 
Combining such a behavior with the factor $\pi^a_{~b}\pi^b_{~a}$ and keeping
in mind that $m_n/4\pi>0$ we have that the factor
$e^{-2\sigma}$ vanishes on the sources, which implies $4 \pi a_n =
m_n$. 
This simplifies the source term in eq.(\ref{eqsigmatilde}). We have  
\begin{equation} e^{-2\tilde \sigma}\sim \frac{8
a_n^2}{\Lambda^2}\left(\frac{z\bar 
z}{\Lambda^2}\right)^{\mu _n-1}~~~{\rm with}~\mu _n=\frac{m_n}{4\pi}.  
\end{equation}
The analogous calculation on the apparent singularites gives a
$e^{-2\sigma}$ regular at those points as expected from eq.(\ref{sigmaeq})
and the equation for $2\tilde\sigma$ takes the simpler form
\begin{equation}
\label{eqsigmatilde2}
2\Delta\tilde\sigma=-e^{-2\tilde\sigma}-4\pi \sum_n
\delta^2(z- z_n)( \mu _n -1)-4\pi\sum_A \delta^2(z- z_A).   
\end{equation}

\subsection{The lapse function}

\bigskip
The variation of the action with respect to $g_{ij}$ \cite{wald,MTW}
gives
$$
\frac{\partial \pi^{ij}}{\partial t}  = 
\frac{Ng^{ij}}{2\sqrt{g}}(\pi^a_{~b}\pi^b_{~a} - (\pi^a_{~a})^2)-
\frac{2N}{\sqrt{g}} 
(\pi^{im}\pi_m^{~j} - \pi^{ij}\pi^a_{~a}) +  \sqrt{g}
(\nabla^i\nabla^jN- g^{ij} \nabla^m \nabla_m  N) 
+
$$
\begin{equation}\label{Neq0}
+ \nabla_m(\pi^{ij}N^m) -\pi^{jm}\nabla_m N^i -\pi^{im}\nabla_m N^j
+N \sum_n \delta^2(x - q_n)
\frac{P_{na}P_{nb}g^{ai}g^{bj}}{2\sqrt{m_n^2+P_{na}P_{nb} g^{ab}}}  
\end{equation} 
where we took into account that in two dimensions
$R^{ij}-g^{ij}R/2=0$, and the variation of the action
with respect to $\pi^{ij}$ gives
\begin{equation}\label{dotg}
\dot g_{ij} = \frac{2N}{\sqrt{g}}(\pi_{ij} - g_{ij}\pi^a_{~a})
+\nabla_j N_i+\nabla_i N_j.
\end{equation}
Using the time evolution given by the above two equations, the relation
$K={\rm const}= 0$ becomes 
\begin{equation}\label{Neq1}
\Delta N= \pi^a_{\ b}\pi^b_{\ a}e^{-2\sigma}N + N
e^{-2\sigma}\sum_n\frac{4P_{nz} P_{n \bar z}}{2 m_n}
\delta^2(z - z_n)
\end{equation}
where we took into account that on the particles
$e^{-2\sigma}$ vanishes.
This is a linear partial differential equation for $N$. We are
interested in the behavior of $N$ near the particle singularity. Near
the particle $n$ the equation takes the form
\begin{equation}\label{Nequation} 
\Delta N = \frac{P_{nz} P_{n \bar z}}{2\pi^2
z \bar z}e^{-2\sigma} N +  \frac {4 P_{nz} P_{n \bar z}}{2 m_n}
e^{-2\sigma} N \delta^2(z). 
\end{equation}
A solution to eq.(\ref{Nequation}) must be such $c = \lim_{r\rightarrow 0} N
e^{-2\sigma}$ is finite. Moreover it is not difficult to prove
that $c =0$. In fact by integrating twice eq.(\ref{Nequation}) in
$|z|$ we find  
\begin{equation}
N \approx \frac{c P_{nz} P_{n\bar z} }{4\pi^2} \ln^2 |z| +c_1 \ln |z| + c_0,
\end{equation}
which substituted in the definition of $c$ gives $c=0$. Moreover as
for $c=0$ the $\delta$ contribution in 
eq.(\ref{Nequation}) is 
absent we have also $c_1=0$. The conclusion is that $N$ is finite
at the particles and the $\delta$ term in eq.(\ref{Neq1}) is absent due to the
vanishing of $Ne^{-2\sigma}$ on the particles and $N$ satisfies the
linear homogeneous equation  
\begin{equation}
\label{eqN2}
\Delta N = e^{-2\tilde\sigma} N.
\end{equation}
Now we analyze the equation 
\begin{equation}
\Delta N = \frac{P_{nz} P_{n\bar z} }{2\pi^2
z \bar z}e^{-2\sigma} N
\end{equation}
for small $z$. Substituting the behavior of $e^{-2\sigma}$ we have
\begin{equation}
\Delta N = \frac{8 \mu _n^2}{\Lambda^2} \left(\frac{z\bar
z}{\Lambda^2}\right)^{\mu _n-1}N.  
\end{equation}
Going over to the variable $r=|z|/\Lambda$ and integrating twice we
have for the $s$-wave around the singularity 
\begin{equation}
\frac{1}{r} \frac{d}{dr}r \frac{d}{dr} N = 8\mu _n^2 (r^2)^{\mu _n-1}N
\end{equation}
and we find
\begin{equation}\label{behaviorofn}
N =  {\rm const.}~[1+ 2 (z\bar z/\Lambda^2)^{\mu _n} + O(z)+ O(\bar z)].
\end{equation}
We examine now the large distance behavior of the
conformal factor. 
Integrating 
\begin{equation}\label{sigmaeq2}
\Delta (2\sigma) =-\pi^a_{\ b}\pi^b_{\ a}e^{-2\sigma}-\sum_n
\delta^2(z -z_n)m_n,  
\end{equation} 
we obtain 
\begin{equation}
\lim_{R\to\infty}2\oint_R \nabla\sigma\cdot \vec n ~ dl= - M \equiv
-4\pi \mu ,
\end{equation} 
where 
\begin{equation}\label{Mintegral} 
M = \sum_nm_n+\int \pi^a_{\ b}\pi^b_{\ a}e^{-2\sigma}\,d^2z\ .
\end{equation} 
which gives for $\sigma$ the asymptotic behavior 
\begin{equation}\label{sigmaabehavior} 
2\sigma\sim - \frac{M}{4\pi}\ln(z\bar z)+\ln s. 
\end{equation} 
This result holds if
the integral appearing in eq.(\ref{Mintegral}) converges. 
As $\sum_n P_n=0$ we have
$\pi^a_{~b} \pi^b_{~a} \approx \frac{1}{r^4}$ and the integral
converges provided $-1 +\mu <0$, which means that the opening
of the cone at infinity 
cannot be negative. (We notice that with a choice $\sum_n P_{nz}\neq 0$ the
integral in eq.(\ref{Mintegral}) would be divergent). Actually
one has infinite solutions to eq.(\ref{sigmaeq2}) depending on
the total energy $M$ of the system which has to be specified in solving
it. We shall see at the end of this section that $M$ cannot depend on time. 

Similarly applying Gauss theorem to
\begin{equation}\label{Neq2} 
\Delta N= e^{-2\tilde\sigma} N = \pi^a_{\ b}\pi^b_{\ a}e^{-2\sigma} N
\end{equation}
we obtain
\begin{equation}\label{Nabehavior}
N \sim \frac{n}{4\pi}\ln(z\bar z)
\end{equation}
with
\begin{equation}\label{nintegral}
n = \int d^2 z \pi^a_{\ b}\pi^b_{\ a}e^{-2\sigma} N.
\end{equation}
The behavior eq.(\ref{sigmaabehavior}) is consistent with the
convergence of the integral 
eq.(\ref{nintegral}) provided $M/4 \pi\equiv \mu <1$ which is
the condition we already 
met. On the other hand  it is easily seen that a behavior $N \sim
|z|^\beta$ which makes the integral (\ref{nintegral}) divergent at infinity is
inconsistent with the differential equation (\ref{Neq2}).

We want now to connect the solution $2\tilde\sigma$ of eq.(\ref{eqsigmatilde2})
with the solutions $N$ of eq.(\ref{eqN2}). We already noticed that
$2\tilde\sigma$ depends on the free parameter $M$, which is the total
energy of the system, and such a parameter does not appear in
eq.(\ref{eqsigmatilde2}). Then it is immediately seen that
\begin{equation}\label{dsigmadM}
N=\frac{\partial (-2\tilde\sigma)}{\partial M}
\end{equation}
is a solution to eq.(\ref{eqN2}). The equation for $N$ being
homogeneous, its normalization is conventional and could also be taken
time dependent. With our choice eq.(\ref{dsigmadM}), the behavior of $N$ at
infinity is, 
from eq.(\ref{sigmaabehavior}) $N\approx \frac{1}{4\pi}\ln(z\bar z)$. On
the other hand it is very simple to prove the following uniqueness
theorem: If $N$ behaves at infinity like $N \approx c_0 + c_1 r^{-a},~
a>0$  and on the particle singularities as in eq.(\ref{behaviorofn}),
then $N\equiv 0$.

In fact multiplying by $N$ eq.(\ref{eqN2})

\begin{equation}
\int_A d^2 z ~\nabla( N\nabla N) = \int_A d^2 z ~\nabla N\cdot \nabla N
+ \int_A d^2 z~ e^{-2\tilde\sigma} N^2.
\end{equation}
But for ${r\rightarrow \infty}$
\begin{equation}
\int_{C_r} N~\nabla N\cdot \vec n ~dl \approx 2\pi (-a)
r^{-a}\rightarrow 0
\end{equation}
and on the particle singularity for ${r\rightarrow }  0$
\begin{equation}
\int_{C_r} N~ \nabla N\cdot \vec n ~dl \approx 2\pi r [c_0 + c_1
(r^2)^{\mu _n}] 
[2 c_1 r \mu _n (r^2)^{\mu _n-1}] \rightarrow 0
\end{equation}
which proves our assertion.
\bigskip

\subsection{The shift functions}
The traceless part of eq.(\ref{dotg}) gives
\begin{equation}
\partial_{\bar z}N^z=- \pi^z_{~\bar z}~e^{-2\sigma} N
\end{equation}
\begin{equation}
\partial_{z}N^{\bar z}=- \pi^{\bar z}_{\ z} ~e^{-2\sigma} N.
\end{equation}
Use eq.(\ref{Neq2}) on the r.h.s. of the first. Then we have
\begin{equation}
2 \pi^{\bar z}_{\ z}(z) \partial_{\bar z} N^z =-4 \partial_{\bar
z}\partial_z N
\end{equation}
whose general solution is
\begin{equation}\label{Nzexp}
N^z =-\frac{2}{\pi^{\bar z}_{\ z}(z)} \partial_z N +g(z)
\end{equation}
where $g(z)$ is an analytic function of $z$. A similar solution holds
for $N^{\bar z}$. The analytic function $g(z)$ must be chosen as to
kill the poles 
generated by the zeros of $\pi^{\bar z}_{\ z}(z)$, which occur only in
presence of three or more particles, and if we are interested in
describing a reference system which 
does not rotate at infinity $g(z)$ has to be chosen as to give $N^z <
|z|$ at infinity. We shall see in sect.7 that,  due to
$\sum_n P_{nz}=0$,  such a function can diverge at infinity at most
linearly. As we 
shall see $g(z)$ encodes the time evolution of all the quantities of
the problem.

The trace part of eq.(\ref{dotg}) gives
\begin{equation}\label{dotsigma}
2\dot\sigma = N^z\partial_z(2\sigma) +\partial_z N^z +{\rm c.c.}
\end{equation}
By substituting the asymptotic behavior for $2\sigma$ eq.(\ref{sigmaabehavior})
into eqs.(\ref{Nzexp},\ref{dotsigma}) we see that $\dot M=0$ as the
logarithmic term $\dot M \ln(z\bar z)/4\pi$ on the l.h.s. is not
matched on the r.h.s.. 

\section{The particle equations of motion}

In the present section we shall derive the general expression for the time
derivative of the particle position $z_n$ and of the particle
conjugate momenta $P_n$.

\noindent
Variation of the action with respect to $P_{nz}$ gives
\begin{equation}\label{dotz}
\dot z_n = -N^i(z_n,t) = - g(z_n).
\end{equation}
The equation for $\dot P_n$ 
\begin{equation}\label{dotP0}
\dot P_{ni}=
\frac{\partial \left(P_{na} N^a-N  
\sqrt{P_{na}P_{nb}g^{ab}+ m_n^2}\right)}{\partial x^i}(z_n)
\end{equation}
requires a little attention because in general the metric is divergent
at the particle position.
Let us consider the expression of $P_n$
\begin{equation}
P_{ni} = m_n \frac{g_{ij}(z_n)(\dot z^j_n + N^j)}{ \sqrt {N^2 -
g_{ij}(\dot z^i_n + N^i )(\dot z^j_n + N^j)}}. 
\end{equation}
At the point $z=z_n$ we have $\dot z^j_n + N^j=0$ but it does not
mean that $P_{ni}=0$ as $g_{ij} = \hat g_{ij}e^{2\sigma}$ is divergent
at that point.
We have to compute the behavior of $N^z(z)-N^z(z_n)$ for $z\rightarrow
z_n$. From eq.(\ref{Nzexp}) we have
\begin{equation}\label{Nzbehav}
N^z(z)-N^z(z_n) \approx \frac{2m_n N(z_n)}{P_{nz}} \left (\frac{(z-z_n)(\bar
z-\bar z_n)}{\Lambda^2}\right)^{\mu _n} +O(z-z_n)+ O(\bar z- \bar z_n) 
\end{equation}
and
\begin{equation}
e^{2\sigma} \approx \frac{P_{nz}P_{n\bar
z}}{m_n^2}\left(\frac{\Lambda^2}{(z-z_n)(\bar z-\bar z_n)}\right)^{-\mu _n} 
+  O(z-z_n)+ O(\bar z- \bar z_n).    
\end{equation}
The square root in the denominator goes over to $N(z_n)$ and thus in
the limit $ z  \rightarrow z_n$ we have
\begin{equation}\label{Pbarz}
m_n \frac{g_{\bar z j}(z_n)(\dot z^j_n + N^j)}{ \sqrt {N^2 -
g_{ij}(\dot z^i_n + N^i )(\dot z^j_n + N^j)}} \rightarrow P_{n\bar z}. 
\end{equation}
With regard to the convergence of the numerator we notice that it is
of the type
\begin{equation}
((z-z_n)(\bar z - \bar z_n))^{-\mu _n} |((z-z_n)(\bar z- \bar
z_n))^{\mu _n} + O(z-z_n) + O(\bar z- \bar z_n)|    
\end{equation}  
which converges absolutely to $1$ only for  $\mu _n <1/2$ i.e. for not
too heavy particle. On the other hand if we average
the l.h.s. of
eq.(\ref{Pbarz}) on a circle around $z_n$ and then take the limit for zero
radius it always converges to $P_{n \bar z}$. We
find a similar problem in writing the equation for $\dot
P_{nz}$. Taking into account the behavior
eqs.(\ref{behaviorofn},\ref{Nzbehav}) of the functions $N$ and $N^z$ for
$z\rightarrow z_n$, eq.(\ref{dotP0}) becomes
\begin{equation}\label{dotP}
\dot P_{n z} = P_{n a}\frac{\partial N^a}{\partial z} -m_n
\frac{\partial N}{\partial z}
\end{equation}  
and a similar equation for $\dot P_{n\bar z}$. Eq.(\ref{dotP}) has to
be understood as taking the average over a circle around $z_n$ and then
taking the limit for zero radius. The procedure is similar to the one
adopted in \cite{ehi}. The situation is analogous to the
one which occurs in electrostatics in presence of point charges. The
electric field is infinite or better not defined on
the particle, but
the average of the electric field on a small sphere surrounding the
particle gives the effective field which acts on the particle as in
such a process the field generated by the particle itself averages to
zero and what survives is the external field. We shall see several
application of such equations in the following.

\section{The two body problem}

\subsection{The conformal factor}

In the two particle case the equation for $\tilde \sigma$ has only two
sources.
\begin{equation} 
\Delta(2\tilde \sigma) = -e^{-2\tilde\sigma} + 4\pi(1-\mu _1)
\delta^2(z-z_1) + 4\pi(1-\mu _2) \delta^2(z-z_2)
\end{equation}
and we have no apparent singularities as $\pi^{\bar z}_{~z}(z)$ has no
zeros in the complex plane. The procedure for solving such equation is
well known \cite{kra}
\begin{equation}\label{fexpression}
e^{-2\tilde\sigma} =\frac{8 f'(z) \bar f'(\bar z)}{(1-f(z)\bar f(\bar z))^2}
\end{equation}
where the function $f(z)$ is given by the ratio of two independent
solutions of the second order fuchsian differential equation with two
regular singularities in $z_1$ and $z_2$ and one at infinity \cite{poole}
\begin{equation} 
f(z) = \frac{k y_1(z)}{y_2(z)}.
\end{equation}
In~ fact~ neglecting~ for~ the~ moment~ the~ sources~ one~ has~
simply~~ to~~ check 
that, being $\ln(f'(z)\bar f'(\bar z))$ an harmonic function, 
\begin{equation} 
4 \partial_f \partial_{\bar f} \ln(1-f\bar f)^2 = -
\frac{8}{(1-f \bar f)^2}.
\end{equation} 
On the singularities the difference of the indices given by \cite{poole}
\begin{equation}\label{index}
y_1\approx(z-z_n)^{\alpha_n},~~~~y_2\approx(z-z_n)^{\beta_n} 
\end{equation} 
has to be
such that on the singularity the term $\ln(f'(z) \bar f'(\bar z))$ gives
\begin{equation} 
4\partial_z\partial_{\bar z}(\alpha_n -\beta_n -1) \ln((z-z_n)(\bar z
- \bar z_n)) = -4\pi(1-\mu _n)\delta^2(z-z_n) 
\end{equation} 
i.e. $\alpha_n -\beta_n = \mu _n$.
The differential equation for $y$ is provided by
\begin{equation}\label{Qequ} 
y''(\zeta) + Q(\zeta) y(\zeta)=0
\end{equation} 
with $\zeta = (z - z_1)/(z_2-z_1)$ and 
\begin{equation}\label{Q2part} 
Q(\zeta) =
\frac{1}{4}\left(\frac{1-\mu _1^2}{\zeta^2}+\frac{1-\mu _2^2}{(\zeta-1)^2}+
\frac{1-\mu _1^2 - \mu _2^2 +\mu _\infty^2}{\zeta(1-\zeta)}\right).
\end{equation}
In fact with two singularities at the finite and one at infinity the
linear residues are fixed by the Fuchs relations in terms of the
$\mu_1$, $\mu_2$ and $\mu_\infty$ (see also eq.(\ref{fuchsrelations})). 
To understand the behavior at infinity let us rewrite
\begin{equation} \label{wronsk}
e^{-2\tilde\sigma} = \frac{8 w_{12}\bar w_{12}}{(y_2\bar y_2 - 
k\bar k ~ y_1 \bar y_1)^2} 
\end{equation}
where $w_{12} = k(y_1' y_2 - y_1 y'_2)$ is the constant wronskian. From
eq.(\ref{wronsk}) we see that at infinity
\begin{equation} 
-2 \tilde \sigma  \approx -(1+\mu _\infty) \ln(\zeta\bar\zeta) 
\end{equation}
or 
\begin{equation} \label{wronsk1}
-2 \tilde \sigma = -2\sigma + \ln(\pi^a_{~b}\pi^b_{~a}) \approx
\frac{M}{4\pi} \ln(\zeta\bar\zeta) - 2 \ln(\zeta\bar\zeta)
\end{equation} 
i.e. $\mu _\infty = 1- M/4\pi \equiv 1 - \mu $.

\noindent
A further requirement on the mapping function $f$ is to give a well
defined, i.e. monodromic $\tilde\sigma$ when one goes around the
singularities $0$, $1$ and consequently $\infty$ in the variable
$\zeta$. From the general theory of fuchsian differential equations we
know that the vector $(k y_1,y_2)$ goes over to a $SL(2,C)$ transformed
vector $(a y_1+b y_2, c y_1+ dy_2)$ due to the constancy of the wronskian.
 From eq.(\ref{wronsk}) 
we see that in order that $e^{-2\tilde\sigma}$ be well defined we need
\begin{equation}
\begin{pmatrix}
a & b \\ c & d
\end{pmatrix} \in SU(1,1). 
\end{equation}
The simplest way to realize this constraint is to choose $f =
ky_1/y_2$ with $y_1$ and $y_2$ behaving as in 
eq.(\ref{index}); then $f$  satisfies the monodromicity condition
around the point $\zeta 
=0$ for any $k$. The point now is to determine $k$ such as the
$\tilde\sigma$ is monodromic also around the point $1$ and as a
consequence at $\infty$. For
completeness we give in the Appendix the expression of $y_1$, $y_2$
and the derivation of $k$.

The so determined value of $k$ gives through eq.(\ref{wronsk}) and
eq.(\ref{sigmatilde}) the conformal factor $e^{2\sigma}$. Then we
obtain $N$ from 
\begin{equation}
N =\frac{\partial (-2\tilde\sigma)}{\partial M}
\end{equation}
whose behavior for large $|z|$ we already know to be $N\approx
\frac{1}{4\pi} \ln(z\bar z)$.

\subsection{Determination of the trajectory}

It is interesting that in order to write down and solve the equations
of motion of 
the particles we do not need to know the explicit form of $2\sigma$
and $N$. In fact from eq.(\ref{dotz}) we see that what we need is $N^z$ in
$z_1$ and $z_2$. For two particles let
\begin{equation} 
2 P = P_1 - P_2;  \quad P= P_1
\end{equation}
from which
\begin{equation} 
\pi^{\bar z}_{z}(z) =  \frac{P_{z}}{2 \pi} \frac{ z_2 -
z_1 }{( z - z_1)( z - z_2)} 
\end{equation}
and thus
\begin{equation}
\pi^a_b\pi^b_a = \frac{2 P_z P_{\bar z}}{4 \pi^2}\frac{(z_1 -
z_2) (\bar z_1 - \bar z_2)}{(z - z_1)(\bar z -
\bar z_1)( z - z_2)(\bar z -\bar z_2)} 
\end{equation}
and
\begin{equation}
N^z = -\frac{4\pi}{P_z}\frac{(z-z_1)(z-z_2)}{z_2-z_1 }\partial_z N + g(z).
\end{equation}
The first term at infinity behaves like
\begin{equation}
-\frac{z}{P_z(z_2-z_1)}
\end{equation}
and in order to have a non rotating reference frame at infinity we
must choose $g(z)$ such as to remove this divergent behavior at infinity
i.e.
\begin{equation}
g(z) = \frac{z}{P_z(z_2-z_1)}.
\end{equation}
As $\pi^{\bar z}_{~ z}$  has poles in $z_1$, $z_2$ we have
\begin{equation}
\dot z_1 = - N^z(z_1) = - g(z_1) = -\frac{z_1}{P_z(z_2-z_1)};~~~
\dot z_2 = -\frac{z_2}{P_z(z_2-z_1)} 
\end{equation}
and then defining $l=z_1-z_2$ we have
\begin{equation}
\dot l =\frac{1}{P_z}.
\end{equation}
We now turn to the equation for $\dot P_z$. 
 We show in general
i.e. for any number of particles that
$\dot P_z$ is given in terms of the residue of the simple poles at the
point $z_n$ of 
the function $Q(z)$ which appears in the fuchsian differential
equation.

Let the singularity at $z_n$ be
 
\begin{equation}
Q(z)= \frac{1-\mu _n^2}{4 z^2}+\frac{\beta_n}{2 z}+  Q_n(z)
\end{equation}
where for the time being $z$ denotes the difference $z-z_n$. The
expansion of the singular solution $y_2$ around the singularity is
\begin{equation}
y_2= z^{\frac{1-\mu _n}{2}}(1+b_n z + O(z^2))
\end{equation}
with
\begin{equation}
b_n= - \frac{\beta_n}{2(1-\mu _n)}.
\end{equation}
Then substituting into eq.(\ref{dotP}) we have in the limit $z\rightarrow
z_n$ 
\begin{equation}\label{dotP2}
\dot P_{n z} = P_{n a}\frac{\partial N^a}{\partial z} -m_n
\frac{\partial N}{\partial z} = 4\pi \frac{\partial \beta_n}{\partial
M}+P_{nz} g'(z_n).
\end{equation}
As we already mentioned, such formulas hold in the general ${\cal N}$-particle
system. In our specific two particle case we have from eq.(\ref{Q2part}) 
\begin{equation}
\beta_1 = -\beta_2 = \frac{1+\mu _\infty^2 -\mu _1^2 -\mu _2^2}{2(z_2-z_1)}
\end{equation}
where working in the $z$ variable the $\beta$'s get divided by
$z_2-z_1$ with respect to the $\beta$'s appearing in
eq.(\ref{Q2part}). Substituting into eq.(\ref{dotP2}) we have
\begin{equation}
\dot P_z = \frac{M}{4\pi(z_2-z_1)}= \frac\mu {z_2-z_1}.
\end{equation}
Thus we have to solve the elementary system
\begin{equation}\label{system}
\dot l =\frac{1}{P_z};~~~~\dot P_z = - \frac\mu {l}.
\end{equation}
Taking the time derivative of the first equation and introducing the
variable $\displaystyle{\lambda = \frac{\ln(l)}{dt}}$ we reach
\begin{equation}
\frac{d\lambda}{\lambda^2}= (\mu -1)dt
\end{equation}
which solved gives
\begin{equation}\label{trajectory}
l = {\rm const}~ [(1-\mu )(t-t_0) - iL/2]^{\frac{1}{1-\mu }}
\end{equation}
which agrees with the solution given in \cite{BCV}.

It is interesting that the same solution can be obtained using simply
the first of eq.(\ref{system}) i.e. $\dot l = 1/P_z$ and the general
conservation laws we shall derive in sect.7.  For two particles the
conserved angular momentum is given by
\begin{equation}
L = \sum_{n=1,2}x^i_n\epsilon_i^{~j} P_{nj} = i(l P_z - \bar l P_{\bar z}).
\end{equation}
On the other hand eq.(\ref{dilatation}) gives for two particles 
\begin{equation}
l P_z + \bar l P_{\bar z} = 2(1-\mu )t + {\rm const}
\end{equation}
which  gives
\begin{equation}
l P_z = (1-\mu ) t - i\frac{L}{2}+ {\rm const}.
\end{equation}
Using now $\dot l=1/P_z$ we reach immediately the solution
eq.(\ref{trajectory}).  

\section{The asymptotic metric}

In the previous section we gave the exact metric for the two particle
case. In case of ${\cal N}$ particles again the solution of the
Liouville equation is given by eq.(\ref{wronsk}) where $y_1,~y_2$ are
solution of a
fuchsian differential equation with singularity at infinity of index
$\mu _\infty = 1 -M/4\pi\equiv 1-\mu $. 
Then the asymptotic behavior of
$\tilde\sigma$ for large $z$ is given by
\begin{equation}
2\tilde\sigma = 2\ln(y_2\bar y_2 - k^2 y_1\bar y_1) - \ln |w_{12}|^2
\approx (2-\mu )\ln(z\bar z) - k^2 (z\bar z)^{\mu -1} +...
\end{equation}
Thus
\begin{equation}\label{asyconf}
e^{2\sigma} \approx{\rm const}~~ (z\bar z)^{-\mu }.
\end{equation}
It follows that the asymptotic behavior of $N$ is
\begin{equation}\label{asyN}
N= \frac{\partial(-2\tilde\sigma)}{\partial M}\approx
\frac{1}{4\pi}\ln(z\bar z) +\frac{k^2}{4\pi} \ln(z\bar z) (z\bar
z)^{\mu -1}+... 
\end{equation}
and then
\begin{equation}
N^z= {\rm const} ~z\left(1+k^2(\mu -1)\ln(z\bar z) (z\bar z)^{\mu -1}+
....\right)+g(z).  
\end{equation}
If $g(z)$ is chosen as to remove the behavior proportional to $z$ 
(non-rotating frame at infinity) the behavior of $N^z$ 
becomes
\begin{equation}\label{asyNz}
N^z \approx  {\rm const} ~z \ln(z\bar z) (z\bar z)^{\mu -1}. 
\end{equation}
Formulas (\ref{asyconf},\ref{asyN},\ref{asyNz}) provide the general
asymptotic behavior of the metric.

In this formalism the appearance of a Gott pair is signalled by an
oscillating behavior of $f(z)$ as $|z| \to \infty$ (the total mass is
now an imaginary number)\cite{BCV}. This pathological behavior is not
surprising because both the formalism and the gauge used here assume
the existence of a global time.

\section{The ${\cal N}$-particle case}

We shall consider in this section the ${\cal N}$-particle case with particular
attention to ${\cal N}=3$. Again the key role is played by the conformal factor
$e^{2\sigma}$ or equivalently $e^{2\tilde \sigma}$ which is given by
eq.(\ref{fexpression}) where $y_1,~y_2$ are solution of a fuchsian
differential equation. It is interesting that $N$ can be rewritten in
the following form
\begin{equation}
N=\frac{\partial G(f)}{\partial f} +\frac{2 \bar f G(f)}{1-f\bar f}+
{\rm c.c.}
\end{equation}
where
\begin{equation}
G(f) = \frac{\partial f}{\partial M}.
\end{equation}
In fact
\begin{equation}
\frac{\partial(-2\tilde\sigma)}{\partial M} =
\frac{1}{f'}\frac{\partial f'}{\partial M} +\frac{2 \bar f
G(f)}{1-f\bar f}+ {\rm c.c.}  
\end{equation}  
and it is easily checked that
\begin{equation}
\frac{1}{f'}\frac{\partial f'}{\partial M}=\frac{\partial}{\partial
f}\left(\frac{\partial f}{\partial M}\right). 
\end{equation}
With regard to the conjugate momenta we have
\begin{equation}
\pi^{\bar z}_{~ z}(z) = \frac{p_{{\cal N}-2}(z)}{\prod_n(z-z_n)}
\end{equation}
where as $\sum_n P_n=0$, $p_{{\cal N}-2}(z)$ is a polynomial of degree ${\cal
N}-2$.  
In flat space the number of Lorentz invariants is easily seen to be
$3 {\cal N}-3$
as for $n>3$ the $2+1$-vector ${\bf P}_n$ is determined by the scalar
product of ${\bf P}_n$ with ${\bf P}_1, {\bf P}_2, {\bf P}_3$. In
curved space we have the same number of invariants. As it is well
known here the invariants are replaced by the traces of the holonomies
around the world lines of an arbitrary collection of particles
\cite{DJH,carrol,menottiseminara}. The
parallel transport monodromy matrices which belong to $SO(2,1)$ or in
the fundamental representation to $SU(1,1)$ are defined up to a
conjugation.  
Keeping in mind that every $SU(1,1)$ holonomy has three real degrees
of freedom we have $3{\cal N}$ degrees of freedom to which we have to
subtract the three degrees of freedom of the $SU(1,1)$ conjugation
thus reaching as expected the same number of invariants as in the flat
case. 

Given at the time $t=0$ the position $z_n$ and the momenta $P_n$
of the particles we know from eq.(\ref{solpiz}) also the position of the
apparent singularities $z_A$. Knowledge of the masses $m_n$ fix also
the second order residues of the poles in $Q(z)$ at $z_n$ to $(1-\mu _n^2)/4$
while the second order residue at the apparent singularities in fixed
to $-3/4$ \cite{yoshida}. 

In order to understand the problem better let us consider the case
${\cal N}=3$. It is useful to go over from the equation in projectively
canonical form 
\begin{equation} 
y''+Q y=0
\end{equation}
where the general form of $Q(z)$ is \cite{yoshida,poole}
\begin{equation}
Q(z) = \sum_n \left[\frac{1-\mu_n}{4(z-z_n)^2} +
\frac{\beta_n}{2(z-z_n)}\right] + \sum_B \left[ -\frac{3}{4(z-z_B)^2}
+\frac{\beta_B}{2(z-z_B)}\right],  
\end{equation} 
to the equivalent form \cite{yoshida}
\begin{equation}\label{equivalent} 
y''+p y' + q y =0
\end{equation} 
with
\begin{equation}\label{Qequivalent}
Q = q- \frac{p'}{2} -\frac{p^2}{4}. 
\end{equation} 
The transformation can be so chosen as to have the following
Riemann scheme
\begin{equation}
\begin{pmatrix}
0&1&z_3&z_A&\infty\\
0&0&0&0&\rho_\infty\\
\mu _1&\mu _2&\mu _3& 2&\rho_\infty+\mu _\infty
\end{pmatrix} 
\end{equation}
where the Fuchs relation 
\begin{equation} 
\mu _1+\mu _2+\mu _3+\mu _\infty +2\rho_\infty=1
\end{equation} 
fixes $\rho_\infty$ in terms of the other parameters and 
\begin{equation} 
p(z) =
\frac{1-\mu _1}{z}+\frac{1-\mu _2}{z-1}+\frac{1-\mu _3}{z-z_3}-\frac{1}
{z-z_A} 
\end{equation} 
\begin{equation} 
q(z) =
\frac{\kappa}{z(z-1)}-\frac{z_3(z_3-1) H }{z(z-1)(z-z_3)}+
\frac{z_A(z_A-1) b_A}{z(z-1)(z-z_A)} 
\end{equation} 
with $\kappa = \rho_\infty(\rho_\infty+\mu _\infty)$.
Here we use the unconstrained parameterization of ref.\cite{yoshida};
the connection with the parameters $\mu_n$ and $\beta_n$ can be easily
obtained by comparing the residues in eq.(\ref{Qequivalent}).

The parameters
$\mu _i$ and $\mu _\infty$ are free real parameters which give the
trace of the holonomies around the particles and infinity (total
energy). $z_A$ is given in terms of the $z_n$ and the $P_n$
($n=1,2,3$) while $H$ is given by the no-logarithm condition i.e. the
requirement that the point $z_A$ is a regular point for
$2\sigma$ \cite{yoshida}. Such $H$ is easily found

\begin{equation}\label{garnierhamiltonian}
H=\frac{z_A(z_A-1)(z_A - z_3)}{z_3(z_3-1)}\left\{b_A^2 - (\frac{\mu _1}{z_A}
+\frac{\mu _2} {z_A-1} +\frac{\mu _3-1}{z_A-z_3})b_A +
\frac{\kappa}{z_A(z_A-1)} \right\}. 
\end{equation} 
Thus we have the following free real parameters
$\mu _1, \mu _2, \mu _3, \mu _\infty, {\rm Re}(z_A),{\rm Im}(z_A), {\rm
Re}(b_A),{\rm Im}(b_A)$ which are eight real parameters, to be
contrasted with the six real 
invariants we have enumerated at the beginning of this section. The
reason is that the constraint that the holonomies described by
eq.(\ref{equivalent}) belong to $SU(1,1)$, fixes the value of ${\rm
Re}(b_A),{\rm 
Im}(b_A)$ in terms of all other parameters. The explicit form of this
relation has been investigated in the mathematical literature
\cite{schlesinger} but we
are not aware of any explicit form of it. Adding a new particle
introduces a new particle singularity with exponent $\mu _4$
i.e. $p(z)$ acquires a new term of the type
\begin{equation} 
\frac{1-\mu _4}{z-z_4}
\end{equation} 
but in addition a new apparent singularity is generated say at
$z_B$. The indices of this new singularity have to be integers,
$0$ and $2$ while the new simple residue $b_B$ has to be a function of
all other parameters, and in particular of $z_B$, because the
$SU(1,1)$ character of the holonomies allow only an increase of $3$ in
the number of free real parameters i.e. $\mu _4, {\rm Re}z_B, {\rm
Im}z_B$, but as we said above we do not know an explicit solution of
this problem. Adding other particles does not alter the procedure.

We come now to the problem of determining the change in time of the
position of the particles $z_n$ and of the position of the apparent
singularities $z_B$. If $f(z) = \frac{k y_1(z)}{y_2(z)}$ provides a
monodromic $2\tilde\sigma$, $N$ is still given by eq.(\ref{dsigmadM}) as the
sources are independent of $M$; the equation for $N^z, N^{\bar z}$
are solved as before by
\begin{equation} 
N^z =-\frac{2}{\pi^{\bar z}_{\ z}(z)} \partial_z N +g(z).
\end{equation} 
The new fact with respect to the two particle case is that now
$\pi^{\bar z}_{~z}(z)$ has ${\cal N}-2$ zeros in the complex plane
which are the position of the apparent singularities. We write
\begin{equation} 
-\frac{\pi^{\bar z}_{~z}(z)}{2}= \frac{1}{4\pi}
\sum_n\frac{P_{nz}}{z-z_n}=\frac{\prod_B(z-z_B)}{{\cal P}(z)}. 
\end{equation} 
Taking the derivative of the previous equation with respect to $z$ and
then setting $z=z_A$ we obtain a relation which will be useful in
the following
\begin{equation}\label{derivativerelation}
\sum_n \frac{P_{nz}}{(z_A-z_n)^2} = -4\pi\frac{\prod_{B\neq
A}(z_A-z_B)}{{\cal P}(z_A)} 
\end{equation}
which for a single apparent singularity reduces to
\begin{equation}
\sum_n \frac{P_{nz}}{(z_A-z_n)^2} = -\frac{4\pi}{{\cal P}(z_A)}. 
\end{equation}
The general expression of $N^z$ is 
\begin{equation}
N^z = \frac{{\cal P}(z)}{\prod_B(z-z_B)}\partial_z N + g(z) 
\end{equation} 
and thus to remove the poles from $N^z$ we must have
\begin{equation}\label{generalg}
g(z) = \sum_B \frac{\partial\beta_B}{\partial M} \frac{1}{z-z_B} \frac{{\cal
P}(z_B)}{\prod _{C\neq B} (z_B-z_C)} + p_1(z), 
\end{equation}
where $p_1(z)$ is a polynomial, as 
\begin{equation}
\partial_z N(z_B) = -\frac{\partial \beta_B}{\partial M}.
\end{equation} 
In fact the expansion of the singular solution around the apparent
singularity $z_A$ is given by
\begin{equation}
y_2=(z-z_A)^{-\frac{1}{2}}(1+\frac{\beta_A}{2}(z-z_A)+O((z-z_A)^2))
\end{equation}
which substituted in $2\tilde\sigma$ gives the above result.  
Recalling the equation of motion of $z_n$ we have now
\begin{equation}\label{zequation}
\dot z_n = - N^z(z_n) = -g(z_n) = -\sum_B
\frac{\partial\beta_B}{\partial M}\frac{1}{z_n-z_B} 
\frac{{\cal P}(z_B)}{\prod _{C\neq
B} (z_B-z_C)} - p_1(z_n). 
\end{equation} 
Thus the time variation of $z_n$ is given in terms of  the momenta $P_n$
which determine completely the polynomial ${\cal P}(z)$ and the derivative
with respect to $M$ of the residues $\beta_A$.

We come now to the equations for $\dot P_n$. We already saw on general
grounds that 
$$
\dot P_{n z} = 4\pi \frac{\partial \beta_n}{\partial M}+P_{nz} ~
g'(z_n)=
$$ 
\begin{equation}
=4\pi \frac{\partial \beta_n}{\partial M} - P_{nz}\sum_B
\frac{\partial \beta_B}{\partial M} \frac{{\cal P}(z_B)}{(z_n-z_B)^2
\prod_{C\neq B}(z_B-z_C)}
+ P_{nz} ~p'_1(z_n).  
\end{equation} 
 From the fuchsianity of the differential equation in projectively canonical
form we know that
\begin{equation}\label{fuchsrelations}
\sum_n\beta_n +\sum_b\beta_B=0;~~~~1-\mu ^2_\infty = \sum_n(1-\mu _n^2
+2\beta_n z_n) + \sum_B (-3 +2\beta_B z_B)
\end{equation} 
so that due to $\mu _\infty = 1-M/4\pi$,
\begin{equation}\label{fuchsian}
\sum_n\frac{\partial \beta_n}{\partial M} +\sum_B\frac{\partial
\beta_B}{\partial M}=0;~~~~
\frac{1}{4\pi}(1 -\frac{M}{4\pi}) = \sum_n \frac{\partial
\beta_n}{\partial M} z_n + 
\sum_B \frac{\partial \beta_B}{\partial M} z_B.   
\end{equation} 
We see that provided $p_1(z)$ is a first degree polynomial,
$p(z)=c_0+c_1 z$, the
consistency of the equation $\sum_n P_n=0$ is 
assured by the the first equation in (\ref{fuchsian}). Similarly one easily
checks that the second relation in eq.(\ref{fuchsian}) provides the
law
\begin{equation}
\frac{d}{dt}\sum_n z_n P_{nz}=\frac{d}{dt}\sum_n \bar z_n \bar P_{nz}=
(1-\frac{M}{4\pi}) 
\end{equation} 
where we have exploited eq.(\ref{derivativerelation}) and $\sum_n
P_{nz}/(z_B-z_n) =0$.
Such equation gives both the law of conservation of angular momentum 
\begin{equation}
\dot L = \frac{d}{dt}  \sum_n i (z_n P_{nz} - \bar z_n \bar P_{nz})
=0
\end{equation} 
which we derived in sect.3 
and the ``generalized conservation law'' 
\begin{equation}\label{dilatation}
\frac{d}{dt}\sum_n (z_n P_{nz} + \bar z_n \bar P_{nz})=2(1-\frac{M}{4\pi})
\end{equation}
which appears to be related to a dilatation transformation \cite{pmds}.
To complete the scheme we must provide the expression of the time
derivative of the position of the apparent singularities $z_A$ and the
time derivatives of the simple residues at such singularities $\beta_A$.
The equation for $\dot z_A$ is easy to obtain. One takes the time
derivative of
\begin{equation}\label{zeros}
0=\sum_n\frac{P_{nz}}{z_A-z_n}
\end{equation} 
which allows to express $\dot z_A$ in terms of $\dot z_n$ and $\dot
P_{nz}$ which are already known.
Using eq.(\ref{zeros}) and eq.(\ref{derivativerelation}) we obtain
$$
0= \frac{\prod_{C\neq A}(z_A-z_C)}{{\cal P}(z_A)}[\dot z_A +c_1
z_A + c_0] + 
\sum_n\frac{\partial\beta_n}{\partial M}\frac{1}{z_A-z_n} +
$$
\begin{equation}
+\sum_{B\neq A}\frac{\partial\beta_B}{\partial M} \frac{1}{z_A-z_B}
\left[\frac{\prod_{C\neq A}(z_A-z_C)}{{\cal P}(z_A)}+
\frac{\prod_{C\neq B}(z_B-z_C)}{{\cal P}(z_B)}\right]. 
\end{equation}
Using the no-log condition the second term in the above equation can
be rewritten as
\begin{equation}
-\beta_A\frac{\partial \beta_A}{\partial M}-\sum_{B\neq
A}\frac{\partial \beta_B}{\partial M}\frac{1}{z_A-z_B} 
\end{equation} 
and thus we reach
\begin{equation}\label{dotzA}
\dot z_A = -c_0 -c_1 z_A +\beta_A \frac{\partial \beta_A}{\partial
M}\frac{{\cal P}(z_A)}{\prod_{C\neq A}(z_A-z_C)}  
-\sum_{B\neq A}\frac{\partial\beta_B}{\partial M} \frac{1}{z_A-z_B}
\frac{{\cal P}(z_B)}{\prod_{C\neq B}(z_B-z_C)}. 
\end{equation} 
In the case of three particles i.e. one apparent singularity, the
above equation reduces to
\begin{equation}\label{dotzA3}
\dot z_A = -c_0 - c_1 z_A +{\cal P}(z_A)\beta_A\frac{\partial
\beta_A}{\partial M}.
\end{equation} 
In order to connect with the mathematical literature we shall choose
the meromorphic function $g(z)$ such as to 
keep the position of particle $1$ at $0$ and of particle $2$ at
$1$. For clarity we shall refer first to the case of three
particles i.e. one apparent singularity, and then give the
generalization to ${\cal N}$ particles. 
Thus
\begin{equation}
g(z) = {\cal P}(z_A)\frac{\partial \beta_A}{\partial
M}\frac{z(z-1)}{(z-z_A)z_A(z_A-1) } = {\cal P}(z_A)\frac{\partial
\beta_A}{\partial M}[ \frac{1}{z-z_A} + \frac{1}{z_A} +
\frac{z}{z_A(z_A-1)}].
\end{equation}
Then substituting in eq.(\ref{dotzA3}) we reach
\begin{equation}
\dot z_A = {\cal P}(z_A)\frac{\partial \beta_A}{\partial M}\left(
\beta_A -\frac{1}{z_A-1}-\frac{1}{z_A}\right).
\end{equation}
If we divide the previous expression by $\dot z_3$ which is simply
given by $-g(z_3)$ we obtain
\begin{equation}
\frac{\partial z_A}{\partial z_3} = \frac{\partial H}{\partial b_A}=
\frac{z_A(z_A-1)(z_A-z_3)}{z_3(z_3-1)}\left[2b_A -\frac{\mu _1}{z_A}-
\frac{\mu _2}{z_A-1}-\frac{\mu _3-1}{z_A-z_3}\right]
\end{equation}
i.e. the first Garnier equation, with hamiltonian
eq.(\ref{garnierhamiltonian}) where we took into account that 
\begin{equation}
\frac{\beta_A}{2} = b_A +\frac{1}{2}\left( \frac{1-\mu _1}{z_A}
+\frac{1-\mu _2}{z_A-1}+ \frac{1-\mu _3}{z_A-z_3}\right). 
\end{equation}
We come now to the more difficult task of finding the time variation
of the simple residues $\beta_B$. To accomplish this we shall exploit
the equation
\begin{equation}\label{dotsigma3}
\frac{d(2\sigma)}{d t} = N^z\partial_z(2\sigma) + \partial_z N^z +
                          N^{\bar z}\partial_{\bar z}(2\sigma) +
                          \partial_{\bar z} N^{\bar z} 
\end{equation}
evaluated in the neighborhood of an apparent singularity. The above
equation embodies part of the information carried by the dynamical
equation for $\dot g_{ij}$.

It is more useful to rewrite equation (\ref{dotsigma3}) in terms of the
reduced conformal factor $2\tilde\sigma = 2\sigma -\ln(2\pi^z_{~\bar
z} \pi^{\bar z}_{~z})$. In order to accomplish this we need the
evolution equation for $\pi^{\bar z}_{~z}$ and
$\pi^z_{~\bar z}$ i.e. for
the traceless part of 
$\pi^a_{~b}$. These are given, outside the particle singularities by 
\begin{equation}\label{dotpi}
\dot \pi^{\bar z}_{~ z}= 2 e^{2\sigma}\partial_z(e^{-2\sigma}\partial_z
N) + 2 \pi^{\bar z}_{~z}\partial_z N^z + N^z\partial_z 
\pi^{\bar z}_{~z}. 
\end{equation}
Using eqs.(\ref{dotsigma},\ref{dotpi}) we reach
\begin{equation}
2 \dot{\tilde\sigma} = g(z)\partial_z (2\tilde\sigma)- g'(z) + {\rm c.c.}
\end{equation}
We now recall the fundamental relation of the theory of fuchsian
differential equation \cite{yoshida,kra}, which can be easily proved
starting from 
eq.(\ref{fexpression}) which holds in the general ${\cal N}$-particle case 
\begin{equation}\label{emt}
Q(z) =
-\frac{1}{2}[\partial^2_z(2\tilde\sigma)+\frac{1}{2}(\partial_z
(2\tilde\sigma))^2]
\end{equation}
where the r.h.s. plays the role of the analytic component of the
energy momentum tensor in two dimensional conformal Liouville theories. Taking
the time derivative of eq.(\ref{emt}) we have
\begin{equation}\label{Qequation}
\dot Q(z) = \frac{1}{2}g'''(z) + 2 g'(z) Q(z) +g(z) Q'(z).
\end{equation}
It is interesting that this equation contains the whole system of the
Garnier equations for the time evolution of the apparent
singularities. If we Laurent-expand $g(z)$ about the apparent singularity
$z_A$ 
\begin{equation}
g(z) = \frac{g_{-1}}{z-z_A} + g_0 +g_1 (z-z_A) +
\frac{g_2}{2}(z-z_A)^2 + O((z-z_A)^2) 
\end{equation}
and
$$
Q(z) =  -\frac{3}{4(z-z_A)^2} +\frac{\beta_A}{2(z-z_A)}+ Q_A(z_A) +
Q_A'(z_A)(z-z_A)+ O((z-z_A)^2) = 
$$
\begin{equation}
= -\frac{3}{4(z-z_A)^2}
+\frac{\beta_A}{2(z-z_A)} - \frac{\beta^2_A}{4} + Q_A'(z_A)(z-z_A)+
O((z-z_A)^2)
\end{equation}
The fourth order poles in eq.(\ref{Qequation}) cancel identically
without producing any 
information; the matching of the residue of the third order pole gives
\begin{equation}\label{thirdorder}
\dot z_A = g_{-1}\beta_A  - g_0
\end{equation}
while the matching of the first order pole gives
\begin{equation}
\dot\beta_A = -2 g_{-1} Q'_A(z_A) + g_1 \beta_A - \frac{3}{2}g_2.
\end{equation}
Eq.(\ref{thirdorder}) reproduces the results obtained in
eq.(\ref{dotzA}), as can be seen by computing $g_{-1}$ and $g_{0}$
form eq.(\ref{generalg}).
As an example we can apply the above equations to the three particle
case. If we keep particle 1 at $0$ an particle $2$ at 1, the $g(z)$
function takes the form
\begin{equation}
g(z) ={\cal P}(z_A)\frac{\partial\beta_A}{\partial M}\left[ 
\frac{1}{z-z_A} +\frac{z-z_A}{z_A(z_A-1)} +\frac{1}{z_A-1}
+\frac{1}{z_A} \right]. 
\end{equation}
Then substituting into eq.(\ref{thirdorder}) we have
\begin{equation}
\dot z_A = {\cal P}(z_A)\frac{\partial\beta_A}{\partial M} \left[ \beta_A
-\frac{1}{z_A-1}-\frac{1}{z_A}\right] = \dot z_3 \frac{\partial
H}{\partial b_A}
\end{equation}
and
\begin{equation}
\dot\beta_A = {\cal P}(z_A)\frac{\partial\beta_A}{\partial M} 
\left[ \beta_A \frac{1}{z_A-1}-2 Q'_A(z_A)\right] = - \dot z_3 \frac{\partial
H}{\partial z_A}.
\end{equation}
These are the Garnier equations with hamiltonian
eq.(\ref{garnierhamiltonian}) once we keep in mind that according to
the general equation eq.(\ref{zequation})
\begin{equation}
\dot z_3 = {\cal P}(z_A)\frac{\partial\beta_A}{\partial
M}\frac{z_3(z_3-1)}{z_A(z_A-1) (z_A-z_3)}.
\end{equation}
It is important to remark that the matching of the residues of
eq.(\ref{Qequation}) on the particle singularities reproduce the equation of
motion $\dot z=-g(z_n)$.

\section{Conclusions}

It is useful to summarize here the basic features of our derivations
which are rather straightforward.
In the instantaneous York gauge $K=0$ the momenta $\pi^a_{~b}$
conjugate to the space metric have only two independent components
$\pi^{\bar z}_{~z}$ ($\pi^{z}_{~\bar z}$) which are meromorphic
(antimeromorphic) functions of $z$ whose residues are the particle
momenta $P_n$. Knowledge of such momenta $\pi^a_{~b}$ allows us to
write a Liouville equation for the reduced conformal factor
$\tilde\sigma$, in which the sources are the particle singularities
i.e. the poles of $\pi^{\bar z}_{~z}$ and the apparent singularities,
i.e. the zeros of $\pi^{\bar z}_{~z}$. The function $N$ is given by
the derivative of 
$\tilde\sigma$ with respect to the total energy while $N^z$ is given
in terms of $N$ by
\begin{equation}
N^z = -\frac{2}{\pi^{\bar z}_{~z}}\partial_z N +g(z),
\end{equation}
where $g(z)$ is a meromorphic function which cancels the polar
singularities of the first term and grows at infinity not faster than
$z$. The linear term in $g(z)$ is fixed if we want to deal with a
reference frame which does not rotate at infinity.

The equation of motion for the particle position are $\dot z_n =
-N^z(z_n) = -g(z_n)$ while those for the particle momenta
\begin{equation}
\dot P_{nz} = P_{na}\partial_z N^a - m_n \partial_z N.
\end{equation}
On the other hand the change in time of the position of the apparent
singularities and their residues are given by the ADM equation for
$\dot{\tilde\sigma}$ computed on the apparent singularities, thus
providing the Garnier hamiltonian system.

In the solution for $N$ and consequently in $g(z)$ and $\dot P_n$
there appears the derivative with respect to the total energy $M$ of
the coefficients of the fuchsian differential equation which underlies
the determination of $\tilde\sigma$. I.e. the change in the $\beta$'s
has to be such as to remain in the $SU(1,1)$ class of differential
equation. It has a very simple solution in the case of two particles
while in general it is obviously related to the Riemann-Hilbert
problem even if it is less specific.

\section{Appendix}

The two solutions of eq.(\ref{Qequ}) which near the origin are a power
multiplied by an analytic function are given by \cite{erdelyi}
\begin{equation}
y_i(\zeta) = \zeta^{\frac{1+\mu _1}{2}}(\zeta-1)^{\frac{1-\mu _2}{2}}
u_i(\zeta)  
\end{equation} 
where
\begin{equation}
u_1=F(a,b;c;\zeta); ~~~~u_2 = \zeta^{1-c}F(a-c+1,b-c+1;2-c;\zeta) 
\end{equation}
and 
\begin{equation} 
a=\frac{1}{2}(1+\mu _1-\mu _2+\mu _\infty);~~~~b=\frac{1}{2}
(1+\mu _1-\mu _2-\mu _\infty);~~~~c=1+\mu _1.
\end{equation} 
Under a circuit around point $1$ they mix as $u_i\rightarrow
B_{i1}u_1+B_{i2}u_2$ and thus the demand that $(ky_1,y_2)$ transforms
under a representation of $SU(1,1)$ imposes
\begin{equation} 
|k^2| = |\frac{B_{21}}{B_{12}}|
\end{equation} 
We have \cite{erdelyi}
\begin{equation} 
B_{12} = -2\pi i~
e^{i\pi(c-a-b)}\frac{\Gamma(c)\Gamma(c-1)}{\Gamma(c-a)\Gamma(c-b)\Gamma(b)
\Gamma(a)}
\end{equation} 
\begin{equation} 
B_{21} = 2\pi i~
e^{i\pi(c-a-b)}\frac{\Gamma(2-c)\Gamma(1-c)}{\Gamma(1-a)\Gamma(1-b)
\Gamma(1+a-c)\Gamma(1+b-c)}
\end{equation} 
which gives
\begin{equation} 
|k^2| = |\Delta(a)\Delta(b)\Delta(1-c)\Delta(2-c)\Delta(c-a)\Delta(c-b)|
\end{equation} 
where as usual $\Delta(x) \equiv \Gamma(x)/\Gamma(1-x)$.
\begin{equation} 
\end{equation}

\section{Acknowledgments}

We are grateful to Davide Fabbri for collaboration in the early stages
of this work. We are also grateful to Marcello Ciafaloni and Stanley
Deser for useful discussions.

\eject

\end{document}